\documentclass[aps,graphicx,showpacs]{revtex4}
\usepackage{graphicx}

\def\ket{\rangle}
\def\<{\langle}
\def\>{\rangle}

\begin{document}
\title{Secure Direct Communication with a Quantum One-Time-Pad}
\author{Fu-Guo Deng$^{1,2}$ and Gui Lu Long$^{1,2,3,4}$\thanks{Corresponding author:gllong@mail.tsinghua.edu.cn} }
\affiliation{$^1$ Department of Physics, Tsinghua University,
Beijing 100084,
China\\
$^2$ Key Laboratory For Quantum Information and Measurements,
Beijing 100084, China\\
$^3$ Center for Atomic and Molecular NanoSciences, Tsinghua
University, Beijing 100084, China\\
$^4$Center For Quantum Information, Tsinghua University, Beijing
100084, China}
\date{\today }

\begin{abstract}
Quantum secure direct communication is the direct communication of
secret messages without first producing a shared secret key. It
maybe used in some urgent circumstances. Here we propose a quantum
secure direct communication protocol using single photons. The
protocol uses batches of single photons prepared randomly in one
of four different states. These single photons serve as a
one-time-pad which are used directly to encode the secret messages
in one communication process. We also show that it is
unconditionally secure. The protocol is feasible with present-day
technique.
\end{abstract}
\pacs{03.67.Hk, 03.65.Ud\\  Published in Physical Review A 69,
052319 (2004)} \maketitle

 Quantum key
distribution(QKD) provides a novel way for two legitimate parties
to establish a common secret key over a long distance. Its
ultimate advantage is its unconditional security, the feat in
cryptography. Combining with the one-time-pad scheme in which the
private key is as long as the messages, secret messages can be
communicated safely from one place to another place.  QKD has
progressed quickly \cite{gisin} since Bennett and Brassard
designed the original QKD protocol \cite{bb84}.

 Recently, a novel concept, quantum secure direct
communication (QSDC) is proposed and
pursued\cite{bekw,bf,dll,cai}. Different from QKD whose object is
to establish a common random key between the two remote parties of
communication, QSDC is to transmit the secret message directly
without first creating a key to encrypt them. In 2002, Beige et
al. presented a QSDC scheme based on single-photon two-qubit
states\cite{bekw}. In this scheme the message can be read after a
transmission of an additional classical information for each
qubit, which is similar to a QKD scheme as each bit of key can
represent one bit of secret message with an additional classical
information, i.e, retaining or flipping the bit value in the key
according to the secret message. Bostr\"{o}m and Felbinger put
forward a Ping-Pong QSDC scheme\cite{bf} using
Einstein-Podolsky-Rosen (EPR) pairs\cite{epr} as quantum
information carriers. It is secure for key distribution, but is
only quasi-secure for direct secret communication if perfect
quantum channel is used. However it is insecure even for QKD if it
is operated in a noisy quantum channel, as shown by
W\'{o}jcik\cite{w}. The Ping-Pong protocol can be modified for
secure QKD by taking into account of the procedures proposed by
W\'{o}jcik\cite{w}. Cai found that the Ping-Pong scheme can be
attacked without eavesdropping\cite{cai2}. Meanwhile, we proposed
a two-step secure QSDC protocol with EPR pairs transmitted in
blocks\cite{dll} by modifying a QKD protocol based on EPR
pairs\cite{ll02}. In Ref.\cite{cai}, Cai modifies the Ping-Pong
scheme by replacing the entangled photons with single photons in
mixed state. However it is unsafe in a noisy channel, and is
vulnerable to the opaque attack\cite{cai3}.

QSDC maybe important in some applications.  For instance when the
transmission time is urgent, or the transmission maybe subject to
the danger of destruction. Furthermore, as the technologies for
quantum information improves, the efficiency of quantum
transmissions maybe greatly increased compared to the low rate
transition in present-day laboratories, then secure direct quantum
communication may well become highly demanded and  become an
elegant means for secret communication.

In this paper, we propose a QSDC scheme that uses single photons
in batches. The states of the single photons themselves serve as a
one-time-pad, and they are encoded with the secret message by two
different unitary operations. The scheme maybe viewed as a
modification of the well-known BB84 QKD scheme. Comparing with
protocols using EPR pairs, this scheme is practical and well
within the present-day technology. All in all, it inherits the
unconditionally security merit of the BB84 QKD scheme, and renders
it an attractive choice in practical applications. Here we first
present our QSDC protocol, then we analyze its security by
reducing it to the BB84 QKD protocol.

The security of QKD is the capability of the users to detect
eavesdropping. If no eavesdropping is detected or the
eavesdropping is negligible, the transmissions are retained, and
after some treatment, a  sequence of secret key is produced.
Otherwise the transmissions are abandoned. However the requirement
for secure direct quantum communication is even higher. In
addition to the ability to detect eavesdropping, the users must
ensure that the secret messages encoded do not leak to
eavesdropper before she is detected. For instance in a noiseless
channel, the Ping-Pong protocol is secure for quantum key
distribution, but is insecure for direct communication as it does
not satisfy the second requirement and some message has already
leaked to Eve, the eavesdropper, before she is detected.

Here we first describe the details of our quantum-one-time-pad
QSDC scheme. Suppose Alice wants to transform a secret message to
Bob. Similar to the BB84 QKD protocol\cite{bb84}, Alice and Bob
use two measuring basis(MB), namely the rectilinear basis, i.e.,
\{ $\left\vert H\right\rangle =\left\vert 0\right\rangle $,
$\left\vert V\right\rangle =\left\vert 1\right\rangle $\}  and the
diagonal basis, i.e., \{$\left\vert u\right\rangle
=\frac{1}{\sqrt{2}}(\left\vert 0\right\rangle +\left\vert
1\right\rangle )$, $\left\vert d\right\rangle
=\frac{1}{\sqrt{2}}(\left\vert 0\right\rangle -\left\vert
1\right\rangle )$\}  where $|H\ket$ and $|V\ket$ are the
horizontal and vertical polarization states respectively.  As in
the BB84 QKD protocol, the $|0\ket$ and $|u\ket$ states represent
the binary value 0, and the $|1\ket$ and $|d\ket$ states represent
binary value 1. For simplicity we call them as the
plus-measuring-basis(Plus-MB) and the
cross-measuring-basis(cross-MB) respectively.  We assume ideal
noiseless channel first. The case with a noisy channel will be
discussed later. The quantum-one-time-pad QSDC protocol contains
two phases:

 (1) The secure doves sending phase. Bob prepares a
batch of polarized single photons and sends the photons to Alice.
Each photon is randomly in one of the four polarization states: $
|H\ket$, $|V\ket$, $|u\ket$ and $|d\ket$. We call this batch of
photons at this stage, the A-batch photon as it goes toward Alice.
 After receiving the batch of photons, Alice and Bob check
eavesdropping by the following procedure. Alice selects randomly a
sufficiently large subset of photons from the A-batch, which we
call the S-batch, and she measures each of them using one of the
two measuring-basis randomly. Alice tells Bob the positions, and
the measuring-basis and the result of the measurement for each of
the sampled photons in the S-batch. With this knowledge, Bob can
determine, through the error rate, whether there is any
eavesdropping. The photons left-over in the A-batch after the
eavesdropping is called the B-batch. Apparently, the B-batch is
the difference set of the A-batch and the S-batch: $B=A-S$.  If
the error rate is high, Bob concludes that the channel is not
secure, and the communication is halted. Otherwise, Alice and Bob
continues to the next phase. This is just like that Bob sends a
batch of doves to Alice for carrying the message back. This is
similar to {\it Ping} in the Ping-Pong protocol, but instead of a
single {\it Ping}, our protocol uses a batch of {\it Ping's}. In
addition to operating in batches, another major difference between
our protocol and the Cai protocol\cite{cai} is that we use four
states, but only the $|H\ket$ and $|u\ket$ states are used in
Cai's protocol at this phase, and this makes the Cai protocol
insecure under an opaque attack\cite{cai3}.

(2) The message coding and doves returning phase. After the
security of the B-batch is completed, Alice encodes each of the
photons in the B-batch with one of the two unitary operations,
$I=\left\vert 0\right\rangle \left\langle 0\right\vert +\left\vert
1\right\rangle \left\langle 1\right\vert $,  and $U=i\sigma
_{y}=\left\vert 0\right\rangle \left\langle 1\right\vert
-\left\vert 1\right\rangle \left\langle 0\right\vert $
respectively, according to the secret message. If the secret
message is 0, then operation $I$ is performed, and if it is a 1
the $i\sigma_y$ operation is performed, the same as that in
Ref.\cite{cai}. The nice feature of the $U$ operation is that it
flips the state in both measuring basis,
\begin{eqnarray}
U\left\vert 0 \right\rangle &=-\left\vert 1\right\rangle,
 & U\left\vert 1\right\rangle  =\left\vert 0\right\rangle,\\
 U\left\vert u\right\rangle & =\left\vert d\right\rangle ,
& U\left\vert d\right\rangle  =-\left\vert u\right\rangle.
\label{o4}
\end{eqnarray}

After encoding the photons in the B-batch, Alice returns them to
Bob. As the A-batch is prepared by Bob,  Bob knows the
measuring-basis, and the original state of each  photon in the
B-batch. He uses the same measuring-basis when he prepared the
photon to measure the photon, and reads out the secret messages
directly. To guarantee the security of the whole communication
process, it is necessary for Alice to use randomly some of the
B-batch photons as checking instances. She will announce publicly
the positions and the coded bit values of these checking photons
after the transmission of a batch is completed. This measure gives
Alice and Bob an estimate whether there is an Eve in the line to
intercept their communication. But Eve can only interrupt their
transmission in this phase and could not get any useful
information about the secret message since Eve can not get any
useful information during the B-batch transmission, in the same
way as in the two-step protocol\cite{dll}.

We now discuss the unconditional security of this
quantum-one-time-pad QSDC scheme. First we notice that the
encoding of secret messages in the second phase(doves returning
phase) is identical to the process in a one-time-pad encryption
where the text is encrypted with a random key as the state of the
photon in the B-batch is completely random. In a one-time-pad
encryption, it is completely safe and no secret messages can be
leaked  even if the cipher-text is intercepted by the
eavesdropper. Here the quantum-one-time-pad QSDC protocol is even
more secure than the classical one-time-pad in the sense that an
eavesdropper can not even intercept the whole cipher-text as the
photons' measuring-basis is chosen randomly. Thus the security of
this QSDC protocol
 depends entirely on the first step when Bob sends the A-batch to Alice.

The process for ensuring a secure A-batch of photons is similar to
that in BB84 QKD protocol. The difference between this protocol
and BB84 QKD is that the B-batch photons are stored, whereas all
the photons are measured one by one in the BB84 QKD scheme. The
security of BB84 QKD is assured by means that Alice and Bob choose
randomly sufficient instances for checking eavesdropping. The
process of the QSDC scheme before Alice encodes her message using
the unitary operation is in fact identical to the BB84 QKD
process.  The BB84 QKD has been proven unconditional secure by
several groups\cite{proof}. The BB84 QKD protocol is secure even
when the channel is noisy. In this way, the process for
establishing a secure quantum channel in this QSDC scheme is
proven unconditionally secure by this observation.

 The Holevo bound states that the
mutual information between Bob and Eve satisfies\cite{nielsen}
\begin{equation}
H(B:E)\leq S(\rho )-\sum\limits_{x}P_{x}S(\rho _{x}),  \label{e1}
\end{equation}%
where $\rho =\sum\limits_{x}P_{x}\rho _{x}$ and $\rho _{x}$ is a
quantum state prepared by Bob with probability $P_{x}$, and
$S(\rho _{i})=-tr(\rho
_{i}\log _{2}\rho _{i})$ is the Von Neumann entropy of state $\rho _{i}$\cite%
{nielsen}. If Bob prepares the four states, $\left\vert
H\right\rangle $, $\left\vert V\right\rangle $, $\left\vert
u\right\rangle $ and $\left\vert d\right\rangle $ symmetrically,
then the binary entropy of states prepared by Bob is
$H(B)=\sum\limits_{x}-P_{x}\log _{2}P_{x}=2$. Thus we have
\begin{equation}
H(B:E)\leq S(\rho )-\sum\limits_{x}P_{x}S(\rho _{x})<H(B).
\label{e4}
\end{equation}
That is to say, Alice and Bob can share a sequence of quantum states securely%
\cite{nielsen}.

The essential difference between this protocol and the Ping-Pong
protocol\cite{bf} and its variant\cite{cai} is that the security
of the quantum channel is analyzed first in a batch after batch
manner and the encoding of the message is done only after the
confirmation of the security of the quantum channel, while in the
Ping-Pong protocols the security check and the encoding of
messages are done concurrently. Since the security of the channel
is insured first in this quantum one-time-pad protocol, Eve can
get nothing even though she monitors the rest of the process of
communication. Another major difference between our protocol and
Cai protocol is the different sets of states in the doves sending
phase. The asymmetry of the $|0\ket$ and $|1\ket$ components in
Cai's protocol makes it insecure under the opaque attack.

\begin{figure}[h]
\begin{center}
\caption{ Implementation of the QSDC with optical delays. CE is
the eavesdropping check; SR represents an optical delay; Switch is
used to control the quantum communication process, if the batch of
photons are safe, the switch is on and the message coding is
performed; CM encodes the secret message, M1 and M2 are two
mirrors for in this simple illustrative set-up.}
\includegraphics[width=14cm,angle=0]{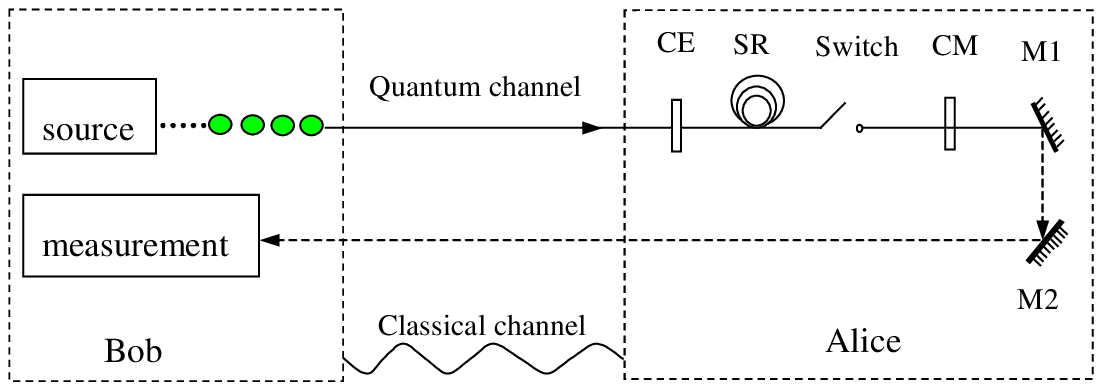} \label{f1}
\end{center}
\end{figure}

In the implementation of this quantum-one-time-pad QSDC protocol,
single photon source and the technique for storing quantum states
are required. These techniques are principally available, for
instance, the single photon source\cite{single}, information
storage through electromagnetic induced transparency\cite{store}.
Of course, they still need further improvement for perfection  for
realistic applications. At present, this protocol can be
implemented with existing techniques. The storage of photons can
be done by optical delays in a fibre as has been proposed in
Ref.\cite{dll}, shown in Fig.{\ref{f1}}. In practice, there are
also losses in the transmission lines, error correcting techniques
are necessary. There have already been quite a few good correcting
codes, for instance, in references\cite{cascade,css,feng}. In
fact, many of the state of art experimental QKD setups use the
Faraday mirrors where the photons are sent to one party and
 then returned back to the sender. It is quite possible that these setups maybe
adapted to realize this QSDC protocol.

Before we conclude, it is worthwhile to inspect the basic
requirements for quantum secure direct communication. First,
eavesdropping check before the message being encoded must be
performed first. This is necessary for secure communication.
Secondly, since eavesdropping can only be performed through
sampling, it is necessary to perform the communications in a batch
after batch manner. A batch of single photons is transmitted first
from one place to another place. Its security is assured by
sampling a sufficiently large subset of instances from the batch.
Then this secured batch is used to encode the secret message and
transmitted to the other location.

Finally, it is seen here that quantum secure direct communication
does not necessarily need EPR pairs as the information carrier,
therefore quantum entanglement and non-locality are not the
necessary requirements for  QSDC.

 This work is supported the
National Fundamental Research Program Grant No. 001CB309308, China
National Natural Science Foundation Grant Nos. 60073009, 10325521,
10244003,  the Hang-Tian Science Fund and the SRFDP program of
Education Ministry of China.

\end{document}